\documentclass{article}
\usepackage{spconf,amsmath,graphicx}
\usepackage{amssymb}
\usepackage{bbm}
\usepackage{xcolor}
\usepackage[T1]{fontenc}
\usepackage{multirow}
\usepackage{graphicx}
\usepackage[vlined,ruled]{algorithm2e}
\SetKwInOut{Require}{Require}

\title{Attention-Guided Adaptation for Code-Switching Speech Recognition}

\name{Bobbi Aditya$^{\star}$\qquad Mahdin Rohmatillah$^{\star}$ \qquad Liang-Hsuan Tai$^{\dagger}$ \qquad Jen-Tzung Chien$^{\star}$}
\address{$^{\star}$Institute of Electrical and Computer Engineering, National Yang Ming Chiao Tung University, Taiwan\\ $^{\dagger}$Electronic and Optoelectronic System Research Labs, Industrial Technology Research Institute, Taiwan}
\begin{document}

\maketitle

\begin{abstract}
The prevalence of the powerful multilingual models, such as Whisper, has significantly advanced the researches on speech recognition. However, these models often struggle with handling the code-switching setting, which is essential in multilingual speech recognition. Recent studies have attempted to address this setting by separating the modules for different languages to ensure distinct latent representations for languages. Some other methods considered the switching mechanism based on language identification. In this study, a new attention-guided adaptation is proposed to conduct parameter-efficient learning for bilingual ASR. This method selects those attention heads in a model which closely express language identities and then guided those heads to be correctly attended with their corresponding languages. The experiments on the Mandarin-English code-switching speech corpus show that the proposed approach achieves a 14.2\% mixed error rate, surpassing state-of-the-art method, where only 5.6\% additional parameters over Whisper are trained.
\end{abstract}

\section{Introduction} \label{sec:intro}
It is common for individuals to use multiple languages for verbal communication in their daily lives \cite{6060895}. A challenging issue in automatic speech recognition (ASR) is to handle the seamless interchange over languages in a speech utterance, known as the code-switching speech \cite{cs-definition}, where the semantics from human may be precisely expressed. The code-switching speech can be inter-sentential or intra-sentential \cite{cs-type}, whereas the intra-sentential case is especially prevalent when a single sentence contains more than one language. To achieve the desirable performance in ASR, it is crucial to cope with the code-switching inputs in an encoder-decoder framework. In recent years, there has been a rapid advance in the development of foundation models for multilingual ASR such as XLS-R \cite{xlsr}, Whisper \cite{Whisper}, USM \cite{USM}, and MMS \cite{MMS}, which have been pre-trained from large corpora of audio data with numerous languages. However, the capability of handling code-switching inputs using these backbone models remains unclear despite their proficiency in capturing inter-sentential speech sentences from multiple languages. In \cite{LM-CodeSwitch}, the limited performance of multilingual language models in code-switching text data was shown. The code-switching speech recognition using multilingual speech models should be also bounded if code-switching scheme is not carefully designed. 

Recently, code-switching ASR \cite{10317410} has obtained significant attention, and can be categorized into two ways through either utilizing a shared encoder to create similar representations or adopting the separate encoders to establish distinct language embeddings. In \cite{oneencoder-3,chien22_interspeech}, a shared encoder was incorporated with language identification (LID) in a training procedure. In \cite{oneencoder-2,leong21_interspeech}, the similarity of token embeddings in a monolingual language was explicitly encouraged to be similar. In addition, the shared encoder \cite{oneencoder-5,9384306,9743719} was extended to develop independent self-attention networks for individual languages. However, in \cite{twoencoder-1,9746154,10191548}, the separate encoders were exploited by utilizing transformer-transducer and CTC model for different languages. While separate encoders improved the recognition results, the additional memory and computation were required.

This paper develops a new approach to code-switching ASR by adapting the pre-trained multilingual ASR model using Whisper based on adapters. A parameter-efficient learning based on an {attention-guided adaptation} is proposed. Initially, a behavior analysis of the attention maps in the transformer decoder is conducted. A head selection method is accordingly proposed to identify those LID token attention heads. These selected heads are then adapted by minimizing an attention-guided loss to ensure the corresponding LID tokens to be correctly attended. In addition, a two-stage adapter training process is designed to carry out a stable parameter-efficient learning \cite{yang23p_interspeech}. The experiments on code-switching Mandarin and English speech recognition show the efficiency and effectiveness using the proposed method.

\begin{figure}[ht] \centering
  {\includegraphics[width=8.5cm]{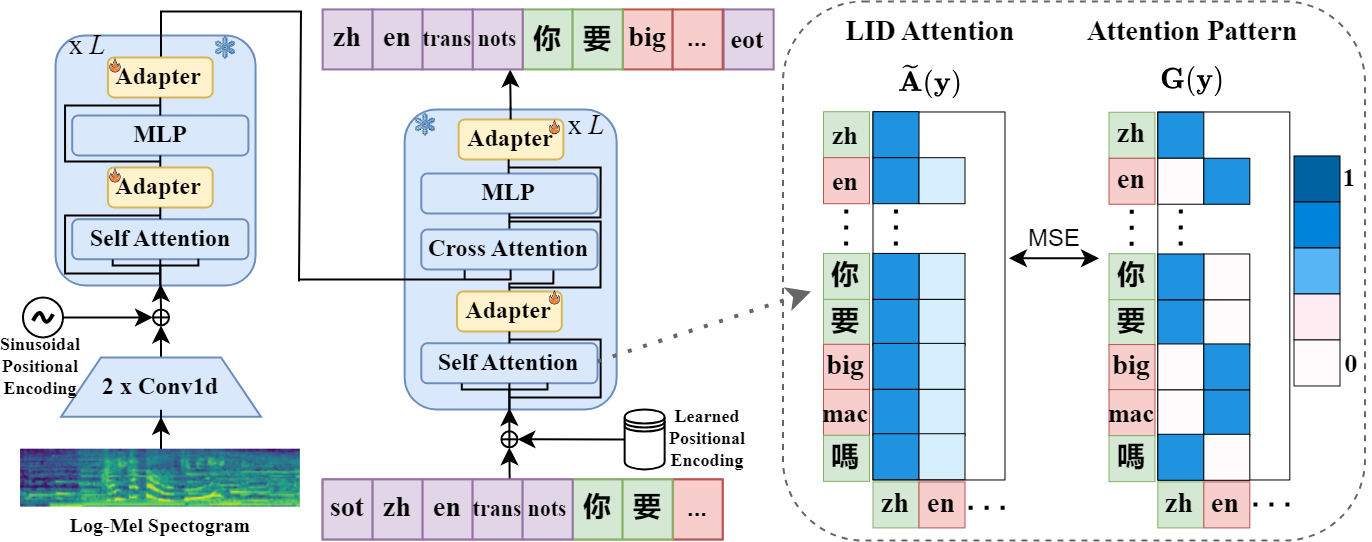}}
  \caption{\label{fig overall}(left) Adapters in encoder and decoder of a transformer. (right) Attention-guided loss calculated from decoder.}
\end{figure}

\section{Code-Switching in Whisper} \label{sec:csWhisper}
Whisper is a state-of-the-art multilingual ASR model with a transformer-based encoder-decoder architecture that has been trained by using 680,000 hours of labeled audio data. This architecture differs from the previous MMS and XLS-R models \cite{MMS,xlsr} that only contained the pre-trained speech encoder. With the inclusion of pre-trained decoder, Whisper presents a notable improvement compared with previous speech foundation models. Given adaptation data $\mathcal{D} = \{\mathbf{x}, \mathbf{y}\}$ where $\mathbf{x}$ is input speech and $\mathbf{y}=\{\mathbf{y}_n\}=\{y_{nm}\}$ is a set of $N$ one-hot vectors in a target transcription, a frozen Whisper backbone $\theta$ combined with the controllable adapter $\phi$ (as shown in Figure \ref{fig overall}) based on an encoder-decoder model $f_{\theta,\phi}$ is constructed by minimizing the cross-entropy loss
\begin{equation} \label{eq ce}
    \mathcal{L}_{\text{CE}}(\mathbf{x},\mathbf{y};\phi)=-\sum_{n=1}^N\sum_{m=1}^M y_{nm}\log p_{nm}
\end{equation}
where $p_{nm}\triangleq p({y}_{nm} | \mathbf{y}_{1}, \ldots, \mathbf{y}_{n-1}, f_{\theta, \phi}(\mathbf{x},\mathbf{y}))$ denotes the decoder output. Each token $\mathbf{y}_n$ is defined under a dictionary of $M$ words. The traditional monolingual transcription is shown with a prompt denoted as $\langle$sot$\rangle$$\langle$lid$\rangle$$\langle$trans$\rangle$$\langle$nots$\rangle$ where $\langle$sot$\rangle$ and $\langle$nots$\rangle$ mean ``start of transcription'' and ``no time stamps'', respectively. Here, $\langle$trans$\rangle$ is used to denote the transcription task for ASR. Now, the bilingual transcription using Whisper has a prompt $\langle$sot$\rangle$$\langle$lid1$\rangle$$\langle$lid2$\rangle$$\langle$trans$\rangle$$\langle$nots$\rangle$ consisting of two LID tokens for Chinese $\langle$zh$\rangle$ and English $\langle$en$\rangle$. A recent work \cite{Whisper-prompt} has extended the utilization of Whisper to handle the task of code-switching ASR by concatenating LID tokens for two languages. The resulting performance was improved by 19\%. However, this work also revealed that Whisper did not accurately identify the language of each word token. The recognition result tended to provide word transcription but refrain from precise language identification. This suggests that the Whisper model unable to handle code-switch settings perfectly.

\begin{figure} \centering
  \includegraphics[width=1.0\linewidth]{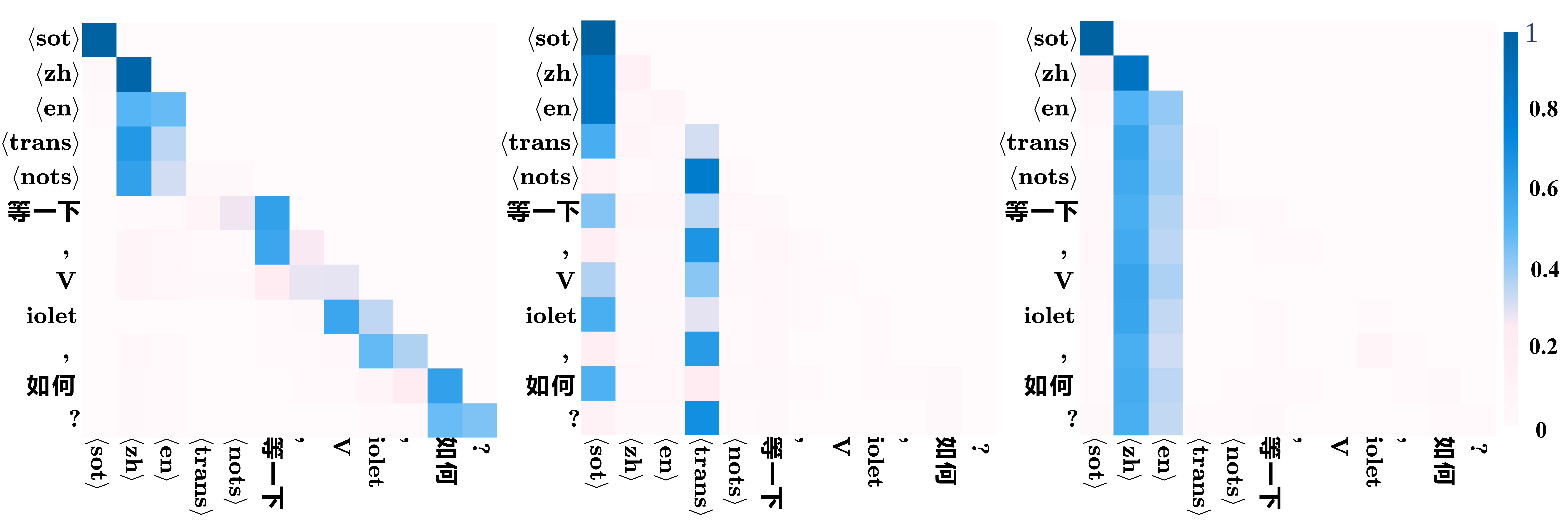}
  \caption{\label{fig:attention_pattern}Examples of three different attention patterns captured by using Whisper model. (left) Self \& neighboring attention obtained from head 5 in layer 4. (middle) Special token attention obtained from head 8 in layer 2. (right) LID token attention obtained from head 11 in layer 8.}
\end{figure}

To investigate how LID tokens behave in a bilingual code-switching ASR task, this study first conducts an insightful analysis on the attention map of each head in a decoder layer. It is interesting to find that there are three typical patterns including self attention, neighboring attention and special token attention \cite{attention-nlp,9306447} as shown by the examples using Whisper model in Figure \ref{fig:attention_pattern} (left and middle subfigures). Notably, a distinctive pattern called the LID token attention is explored as shown in the right subfigure. This paper is accordingly motivated to guide the model learning through minimizing an attention-guided loss with the goal of highlighting the attention behaviors in code-switching speech so that the bilingual ASR decoder is allowed to sensitively pay attention on the language changes in speech signal for word prediction.

\section{Attention-Guided Adaptation} \label{sec:methods}
First of all, the self attention map of a head in transformer decoder is calculated from a target sequence $\mathbf{y}$ via
\begin{equation}
    \mathbf{A}(\mathbf{y})=\operatorname{softmax}\left(\frac{\mathbf{Q} \mathbf{K}^{\top}}{\sqrt{d}}\right) \in \mathbb{R}^{N \times N},
\end{equation}
where $\mathbf{Q}$ and $\mathbf{K} \in \mathbb{R}^{N \times d}$ denote the query and key matrices, respectively, and $d$ is the dimension of keys or queries. As seen in right-hand-side of Figure \ref{fig overall}, the attention-guided adaptation is implemented by first selecting the attention maps in decoder reflecting the LID and then learning the adapters to come out with self attention maps to align LID tokens.

\subsection{Attention Head Selection} \label{ssec:headselection}
It is crucial to select the attention heads that concentrate on LID information in code-switching speech. These heads are then guided via adapter learning in bilingual ASR where the attention-guided loss is minimized. Let $\{\mathbf{A}_{\ell h}\}$ denote the individual attention maps from $L$ layers where each layer has $H$ heads. Notably, backbone $\theta$ is required to calculate $\mathbf{A}_{\ell h}$. To identify those LID attention maps $\mathbf{A}_{\ell h}$, or simply denoted by an $N$-dimensional matrix $\mathbf{A}=[A_{ij}]_{N \times N}$ with entry value $A_{ij}(\mathbf{y})$ for self attention between tokens $i$ and $j$ calculated from $\mathbf{y}$, this paper introduces the language indicator function
\begin{equation}
\mathbb{I}(\mathbf{A}) = 
\begin{cases}
1, & \text{if} \; \sum_{i=1}^{N} \sum_{j \in \Omega} A_{ij}(\mathbf{y}) > \sum_{i=1}^{N}\sum_{k \in \overline{\Omega}} A_{ik}(\mathbf{y}) \\
0, & \text{otherwise}
\end{cases}
\end{equation}
where $\Omega = \{\langle$zh$\rangle, \langle$en$\rangle\}$ denotes the LID tokens and $\overline{\Omega}$ denotes the other $N-2$ non-LID tokens. The set of LID attention maps $\widetilde{\mathbf{A}}$ is then formed with $\mathbb{I}(\mathbf{A})=1$ over $\mathbf{y} \in \mathcal{D}$. The heads selected with top $K$ highest counts are found by
\begin{equation} \label{eq lid}
\widetilde{\mathbf{A}}(\mathbf{y}) = \{ \mathbf{A}_{\ell h} \vert (\ell,h) \in \textstyle\mathop{\mathrm{arg\,top}\,k}_{(\ell, h)} \textstyle\sum_{\mathbf{y} \in \mathcal{D}} \mathbb{I}(\mathbf{A}_{\ell h}) \}
\end{equation}
through a forward pass over all target sentences $\mathbf{y} \in \mathcal{D}$.

\subsection{Attention-Guided Adaptation} \label{ssec:attguided}
Next, the attention-guided adaptation is performed via adapter learning where the resulting LID attention maps $\widetilde{\mathbf{A}}(\mathbf{y})$ are guided towards their ground-truth attention maps $\mathbf{G}(\mathbf{y})=[G_{ij}(\mathbf{y})]_{N\times N}$. As seen in the right-hand-side of Figures \ref{fig overall} and \ref{fig:attention_pattern}, those LID attention maps with high attention values are guided to be \textit{aligned} with the correct LID tokens in $\Omega = \{\langle$zh$\rangle, \langle$en$\rangle\}$ which are provided in the target $\mathbf{y}$. The attention pattern as a kind of code-switching map is specified as
\begin{equation}
G_{i,j\in \Omega}(\mathbf{y}) = \begin{cases}
c & \text{if } \text{token}~i~\in \langle\text{zh}\rangle/\langle\text{en}\rangle, j \in \langle\text{zh}\rangle/\langle\text{en}\rangle\\
0 & \text{otherwise}
\end{cases}
\end{equation}
Here, only the columns $j$ corresponding to LID tokens $\langle$zh$\rangle$ and $\langle$en$\rangle$ are guided. Since the values of the other columns are kept unchanged, a soft label with constant $c$ is assigned for LID token in a range $0.5<c<1$ so as to preserve non-LID information in the other tokens during guidance learning. The attention-guided loss, controlled by decoder $\phi_d$ and calculated over target data $\mathbf{y}$, is therefore yielded by
\begin{equation} \label{eq ag}
\mathcal{L}_{\text{AG}}(\mathbf{y}; \phi_d) = \sum_{\mathbf{A} \in \widetilde{\mathbf{A}}}\sum_{i=1}^N \sum_{j \in \Omega} \left(A_{ij}(\mathbf{y},\phi_d)-G_{ij}(\mathbf{y})\right)^2.
\end{equation}
Notably, the regression errors over training samples $\mathbf{y}$ between attention maps and code-switching maps are accumulated over all LID attention heads $\mathbf{A}_{\ell h}$ selected in $\widetilde{\mathbf{A}}(\mathbf{y})$ where all of attention entries over the columns of LID tokens $\langle$zh$\rangle$ and $\langle$en$\rangle$ are counted. This loss is minimized to guide the learned decoder to carefully attend the language changes during word prediction. This guidance is only performed for LID tokens. The attention maps corresponding to non-LID tokens are unchanged to preserve all non-LID information in ASR.

\subsection{Two-Stage Optimization Procedure} \label{ssec:petraining}
This study accordingly carries out an efficient adaptation where the adapters \cite{houlsby} in encoder $\phi_e$ and decoder $\phi_d$ are learned with the frozen backbone transformer $\theta$. The attention guidance (AG) is performed through the optimization over a classification problem
\begin{equation}
\phi_e, \phi_d = \underset{\phi_e, \phi_d}{\arg\min}~E_{\mathcal{D}} \left[ \mathcal{L}_{\text{CE}}(\mathbf{x}, \mathbf{y};\phi) + \gamma \mathcal{L}_{\text{AG}}(\mathbf{y};\phi_d) \right]
\end{equation}
where $\phi=\{\phi_e,\phi_d\}$ and $\gamma$ denotes the hyperparameter to balance between joint training over two objectives. A two-stage optimization procedure is implemented in Algorithm \ref{algo:methhod}. Instead of directly estimating encoder $\phi_e$ and decoder $\phi_d$ in a hybrid model, this paper presents a stable learning consisting of two stages. The first stage is to estimate adapter encoder $\phi_e$ by minimizing $\mathcal{L}_{\text{CE}}$ where the backbone Whisper $\theta$ is utilized by disregarding adapter decoder $\phi_d$. Given a converged encoder, the second stage is to jointly train $\phi_e$ and $\phi_d$ by minimizing both classification loss $\mathcal{L}_{\text{CE}}$ and guidance loss $\mathcal{L}_{\text{AG}}$ where backbone $\theta$ is again utilized. The attention guidance via $\mathcal{L}_{\text{AG}}$ acts as an auxiliary objective to constrain adapter learning to handle code-switching ASR by minimizing $\mathcal{L}_{\text{CE}}$.

\begin{algorithm}[ht]
\DontPrintSemicolon
\SetAlgoLined
\caption{\label{algo:methhod}Attention-guided adapter learning}
\Require{dataset $\mathcal{D}=\{\mathbf{x}, \mathbf{y}\}$, backbone $\theta$, adapter encoder $\phi_e$, decoder $\phi_d$, weight $\gamma$}
initialize $\phi_e$, $\phi_d$ and fix $\theta$\\
select attention maps $\widetilde{\mathbf{A}}$ in (\ref{eq lid}) by forward pass over $\mathcal{D}$ \\
\textbf{stage 1:} update encoder $\phi_e$\\
 \For{each batch $(\mathbf{x},\mathbf{y})$ in $\mathcal{D}$}{
        compute $\mathcal{L} = \mathcal{L}_{\text{CE}}(\mathbf{x}, \mathbf{y};\phi_e)$ in (\ref{eq ce})\\
        update $\phi_e$ using $\nabla_{\phi_e}\mathcal{L}$\\
     }
\textbf{stage 2:} update encoder $\phi_e$ and decoder $\phi_d$\\
    \For{each batch $(\mathbf{x},\mathbf{y})$ in $\mathcal{D}$}{
        compute $\mathcal{L}_{\text{AG}}( \mathbf{y};\phi_d)$ in (\ref{eq ag}) \& $\mathcal{L}_{\text{CE}}( \mathbf{x},\mathbf{y};\phi)$ in (\ref{eq ce}) \\
        integrate $\mathcal{L} = \mathcal{L}_{\text{CE}}(\mathbf{x}, \mathbf{y};\phi_e,\phi_d) + \gamma \cdot \mathcal{L}_{\text{AG}}( \mathbf{y};\phi_d)$\\
        update $\phi_e$\& $\phi_d$ using $\nabla_{\phi_e}\mathcal{L}$ \& $\nabla_{\phi_d}\mathcal{L}$\\
     }
\end{algorithm}
\section{Experiments} \label{sec:experiment}
The experiments were carried out under an end-to-end speech processing using ESPnet Toolkit \cite{espnet}. A Mandarin-English code-switching speech corpus in South-East Asia (SEAME) \cite{seame} from Singaporean and Malaysian speakers was used.

\subsection{Experimental Settings} \label{ssec:expsetup}
The multilingual Whisper-small with $H=12$ heads and $L=12$ transformer layers in encoder and decoder was used. The adapter hidden dimension size was $192$, with $\gamma=0.01$ and $c=0.6$ were chosen. Following the original Whisper training process, AdamW optimizer with learning rate $1\times10^{-3}$ was used. For each training stage, the model was trained by 15 epochs with the final model which was obtained by averaging the weights of three best models, selected based on validation loss from the best epoch. To evaluate the performance of code-switching ASR, this paper made comparison over three settings including (a) one-stage adapter, (b) one-stage adapter with the proposed attention guidance (w/ AG), (c) two-stage adapter w/ AG. For ablation study on attention head selection, the results were carried out by selecting different attention heads during AG loss calculation based on two-stage adapter w/ AG setting. In the evaluation, the devman set (dominated by Mandarin speech) and devsge set (dominated by Singaporean English speech) of SEAME were used. In each set of experiments, separate measures on monolingual and code-switching utterances were reported. The performance metrics included word error rate (WER), character error rate (CER), and mixed error rate (MER) for English, Chinese, and code-switching speech, respectively. Parameter size was compared.

\begin{table}[th]
\caption{\label{table:AGresult}Comparison of different methods for code-switching ASR on devman and devsge of SEAME dataset.}
\resizebox{\columnwidth}{!}{%
\begin{tabular}{ccccccc}
\hline
Method                                                                            & Dev Set & \multicolumn{1}{l}{WER} & \multicolumn{1}{l}{CER} & \multicolumn{1}{l}{MER} & \begin{tabular}[c]{@{}c@{}}Overall \\ MER\end{tabular} & \begin{tabular}[c]{@{}c@{}}Train\\ Param\end{tabular}              \\ \hline
\multirow{2}{*}{\begin{tabular}[c]{@{}c@{}}Original \\ Prompt \cite{Whisper-prompt}\end{tabular}}          & devman  & 76.3                    & 24.7                    & 38.2                    & 38.2                                                   & \multirow{2}{*}{--}                                                          \\
                                                                                  & devsge  & 82.8                    & 32.4                    & 56.4                    & 65.0                                                   &                                                                             \\ \hline
\multirow{2}{*}{\begin{tabular}[c]{@{}c@{}}Modified \\ Prompt \cite{Whisper-prompt}\end{tabular}}   & devman  & 45.8                    & 23.6                    & 33.4                    & 32.7                                                   & \multirow{2}{*}{--}                                                          \\
                                                                                  & devsge  & 46.7                    & 31.0                    & 49.6                    & 47.6                                                   &                                                                             \\ \hline
\multirow{2}{*}{SOTA \cite{espnet,Whisper-prompt}}                                                        & devman  & --                       & --                       & --                       & 16.6                                                   & \multirow{2}{*}{47.3 M}                                                    \\
                                                                                  & devsge  & --                       & --                       & --                       & 23.3                                                   &                                                                             \\ \hline
\multirow{2}{*}{\begin{tabular}[c]{@{}c@{}}One-Stage\\Adapter\end{tabular}} & devman  & 25.0                    & 15.6                    & 14.4                    & 15.1                                                   & \multirow{2}{*}{\begin{tabular}[c]{@{}c@{}}14.3 M \\ (5.6\%)\end{tabular}} \\
                                                                                  & devsge  & 24.7                    & 20.4                    & 19.7                    & 21.6                                                   &                                                                             \\ \hline
\multirow{2}{*}{\begin{tabular}[c]{@{}c@{}}One-Stage\\Adapter w/ AG\end{tabular}} & devman  & 22.0                    & 15.5                    & 13.8                    & 14.4                                                   & \multirow{2}{*}{\begin{tabular}[c]{@{}c@{}}14.3 M \\ (5.6\%)\end{tabular}} \\
                                                                                  & devsge  & 24.2                    & 19.1                    & 19.8                    & 21.3                                                   &                                                                             \\ \hline
\multirow{2}{*}{\begin{tabular}[c]{@{}c@{}}Two-Stage\\Adapter w/ AG\end{tabular}} & devman  & 22.2                    & 14.9                    & 13.5                    & 14.2                                                   & \multirow{2}{*}{\begin{tabular}[c]{@{}c@{}}14.3 M \\ (5.6\%)\end{tabular}} \\
                                                                                  & devsge  & 24.2                    & 19.4                    & 18.9                    & 20.8                                                   &                                                                             \\ \hline
\end{tabular}}
\vspace*{-\baselineskip}
\end{table}

\subsection{Experimental Results} \label{ssec:results}
Table \ref{table:AGresult} shows the results of various code-switching ASR approaches. The baseline performance incorporates the findings in \cite{Whisper-prompt} which include original and modified prompt using Whisper. The state-of-the-art (SOTA) result in \cite{espnet,Whisper-prompt} that followed the ESPnet SEAME recipe is provided. The result on one-stage adapter training is listed. The results on one-stage adapter w/ AG clearly demonstrate the efficacy of introducing AG loss, leading to an overall improvement compared with one-stage adapter training w/o AG. Furthermore, introducing two-stage training scheme enhances the results, as indicated by the overall MER. Notably, two-stage adapter w/ AG outperforms the methods in \cite{Whisper-prompt, espnet} across all metrics while requiring much fewer trainable parameters. Compared with backbone Whisper, AG only learns 5.6\% of parameters.

\begin{table}[th]
\vspace*{-\baselineskip}
\caption{\label{table:HSresult}Ablation study on attention head selection.}
\resizebox{\columnwidth}{!}{%
\begin{tabular}{cccccc}
\hline
\begin{tabular}[c]{@{}c@{}}Head Selection\end{tabular}               & \multicolumn{1}{l}{Dev Set} & \multicolumn{1}{l}{WER} & \multicolumn{1}{l}{CER} & \multicolumn{1}{l}{MER} & \begin{tabular}[c]{@{}c@{}}Overall \\ MER\end{tabular} \\ \hline
\multirow{2}{*}{\begin{tabular}[c]{@{}c@{}}Random  50\% of \\ $\{\mathbf{A}_{\ell h}\}$\end{tabular}} & devman                      &  75.2                       &     58.8                    &   61.2                      &   60.7                                                     \\
                                                                        & devsge                      &  81.2                       &  78.7                       &  64.5                       &  78.8                                                      \\ \hline
\multirow{2}{*}{\begin{tabular}[c]{@{}c@{}}$\widetilde{\mathbf{A}}$ (100\%)\end{tabular}}      & devman                      & 21.5                    & 14.8                    & 13.5                    & 14.1                                                   \\
                                                                        & devsge                      & 25.4                    & 18.7                    & 18.7                    & 21.1                                                   \\ \hline
\multirow{2}{*}{\begin{tabular}[c]{@{}c@{}} $\widetilde{\mathbf{A}}$ (top 60\%)\end{tabular}}  & devman                      & 22.2                    & 14.9                    & 13.5                    & 14.2                                                   \\
                                                                        & devsge                      & 24.2                    & 19.4                    & 18.9                    & 20.8                                                   \\ \hline

\end{tabular}}
\end{table}

\begin{figure}[th]
\begin{minipage}[b]{1.0\linewidth}
  \centering
  \centerline{\includegraphics[width=1.0\linewidth]{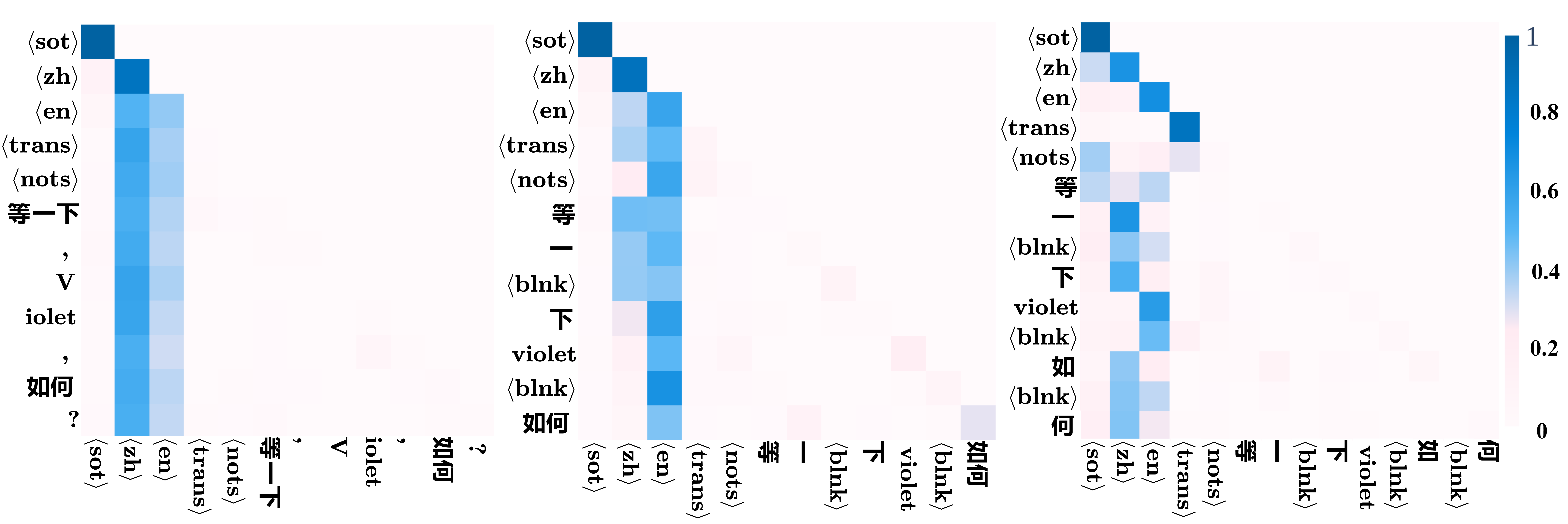}}
  \caption{LID token attention from head 8 in layer 2 using three different models: (left) backbone w/o adapter, (middle) one-stage adapter, and (right) two-stage adapter w/ AG. This comparison shows the attention-guided models correctly attends LID of each word token. \text{$\langle$blnk$\rangle$} stands for blank token.}
  \label{fig:AGanalysis}
\end{minipage}
\end{figure}

The outcome of table \ref{table:HSresult} shows the importance of identifying attention heads for LID tokens. Selecting random heads would lead to a failure of model convergence due to the misguided attention. In contrast, when LID token attention heads are chosen, the model converges and achieves the best performance. However, given that a significant portion of the model's heads are language-related, where $\mathbb{I}(\mathbf{A}_{\ell h})=1$ (110 out of 144 heads in the study), it becomes vital to further reduce the number of selected heads. Typically, selecting most of LID heads results in an overly aggressive guidance process, causing the model to lose information beyond language. Consequently, the best result is obtained when guiding only 60\% of the selected LID token attention heads.

\subsection{Experimental Analysis} \label{ssec:Analysis}
The comparison of LID token attention maps of original Whisper w/o adapter, one-stage adapter and two-stage adapter w/ AG can be seen in Figure \ref{fig:AGanalysis}. In the absence of adaptation process, the attention map shows that all tokens in a sequence attend both LID tokens, regardless of the language of word token. If the model is adapted without the AG process, the attention map shows that the word tokens attends the wrong LID token. In contrast, the model trained with AG generates an attention map that closely aligns with the ground-truth attention pattern. Namely, each word token correctly attends to its corresponding LID token. This finding highlights the significant impact of AG process where each word token is guided to capture language changes during word prediction.

During the attention head selection process, it is found that a majority of attention heads in the model are categorized as LID token heads where $\mathbb{I}(\mathbf{A}_{\ell h})=1$. Naturally, multiple head attention in transformers will capture different kinds of relation in a sequence. This discovery underscores the presence of noteworthy redundancy within the model, where multiple heads are capturing similar information, which may deviate from typical behavior of transformers. It is believed that these findings warrant further investigation. 

\section{Conclusions} \label{sec:conclusion}
This study proposes an attention-guided adaptation for code-switching ASR by adapting the Whisper with two-stage adapter training. By analyzing the attention maps in the transformer decoder of Whisper, a distinctive LID token attention is discovered. A head selection process is then applied to identify those LID token attention heads, along with the attention-guided process to ensure correct language identification and guide the ASR decoder to attend the code-switching in speech signal. Experiments on code-switching Mandarin and English speech recognition using SEAME dataset demonstrate the efficiency and effectiveness of the proposed method by improving the overall MER of previous SOTA results while using the reduced trainable parameters.

\bibliographystyle{IEEEbib}
\small
\bibliography{refs}

\end{document}